\documentclass[prl,twocolumn,,superscriptaddress,showpacs]{revtex4}%

\usepackage{amsfonts}%
\usepackage{amsmath}%
\usepackage{amssymb}%
\usepackage{graphicx}%

\begin{document}

\title{Topological Hysteresis in the Intermediate State of Type-I Superconductors}

\author{Ruslan Prozorov}%
\affiliation{Ames Laboratory and Department of Physics \& Astronomy, Iowa State University, Ames,
Iowa 50011, U.S.A.}

\author{Russell W. Giannetta}%
\affiliation{Loomis Laboratory of Physics, University of Illinois at Urbana - Champaign, 1110 West
Green Street, Urbana, IL 61801, U.S.A.}

\author{Anatolii A. Polyanskii}%
\affiliation{Applied Superconductivity Center, University of Wisconsin, 1500 Engineering Drive,
Madison, WI 53706, U.S.A.}

\author{Garry K. Perkins}%
\affiliation{Blackett Laboratory, Imperial College of Science
Technology and Medicine, London SW7 2BZ, United Kingdom}



\begin{abstract}
Magneto-optical imaging of thick stress-free lead samples reveals two distinct topologies of the
intermediate state. Flux tubes are formed upon magnetic field penetration (closed topology) and
laminar patterns appear upon flux exit (open topology). Two-dimensional distributions of shielding
currents were obtained by applying an efficient inversion scheme. Quantitative analysis of the
magnetic induction distribution and correlation with magnetization measurements indicate that
observed topological differences between the two phases are responsible for experimentally
observable magnetic hysteresis.
\end{abstract}

\date{7 July 2005}

\maketitle

The structure of the intermediate state in type-I superconductors has a long history beginning with
the pioneering papers of Landau \cite{1,2} and continuing to the present day
\cite{3,4,5,6,7,8,9,10,11,12}. Intermediate state flux patterns closely resemble those found in a
wide variety of hydrodynamic, chemical and solid state systems \cite{9,9a,9b}. Study of the
intermediate state is therefore vital to a general understanding of pattern formation. Flux
structures can be readily tuned with a magnetic field and imaged with magneto-optical (MO)
techniques \cite{3,4}. The observed patterns can then be correlated with underlying thermal,
magnetic and resistive properties. Early MO images of the intermediate state revealed a variety of
phenomena not predicted by the simple theory \cite{3,4}. The initial models were then refined to
include domain branching and corrugation. Still, it is widely believed that a true
thermodynamically stable configuration of the intermediate state is the famous Landau laminar
structure \cite{2}.

Magnetic hysteresis is routinely observed in type-I superconductors and has generally been
attributed to impurities, grain boundaries, dislocations and other imperfections of the crystal
structure \cite{4}. In this Letter we focus on the relationship between the topology of flux
structures and their macroscopic magnetic properties. We find that a small residual hysteresis
remains even in the most carefully prepared samples and present evidence that this hysteresis
arises from the differences in the topology of the intermediate state between flux entry and flux
exit.

\begin{figure}[tb]
\centerline{\includegraphics[width=8cm]{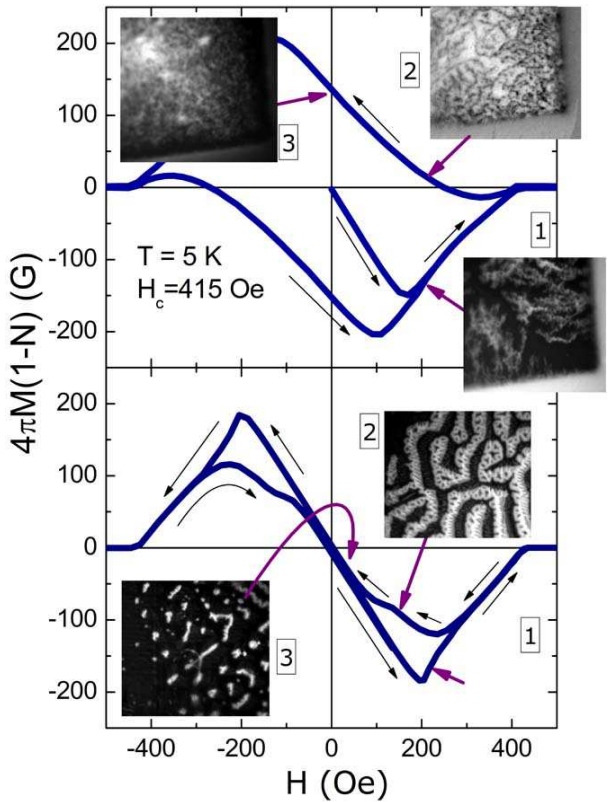}}%
\caption{Magnetization loops in stressed (top) and stress-free sample (bottom). Also shown MO
images obtained at magnetic fields indicated by arrows (
the numbers indicate images in succession).}%
\label{fig1}%
\end{figure}

Samples were prepared from 99.9999{\%} lead \footnote{various sources, including
\emph{Puratronic}$^\circledR$ from \emph{Alfa Aesar}} foils and rods. More than a dozen
samples were prepared by using various annealing protocols or deliberately introducing
stress by cold rolling. The most reversible samples were obtained by melting lead between
two \textit{Pyrex} slides. Samples had thickness between $d=$ 0.1 and 1.5 mm and were
about $1.5\times 1.5\;\mathrm{mm}^\mathrm{2}$ in planar dimension. The topological
features described here were thickness independent above $d\approx 0.5\,\mathrm{mm}$,
which indicates that they are not due to surface-related effects. We show data for
samples which had demagnetization factors of about $N=0.5$ (determined both from initial
magnetisation and direct calculations \cite{20}.) \textit{Quantum Design} MPMS
magnetometer was used for DC magnetization measurements. MO imaging was performed in a
pumped flow-type optical $^{4}$He cryostat using Faraday rotation of polarized light in
Bi-doped iron-garnet films with in-plane magnetization \cite{14}. In all images the
bright regions correspond to the normal state and dark regions to the superconducting
state.

\begin{figure}[tb]
\centerline{\includegraphics[width=8cm]{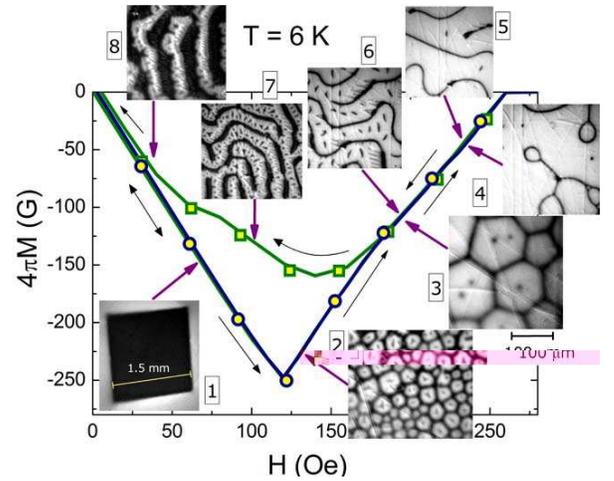}}%
\caption{Magnetization loop in a stress-free sample accompanied by MO and zfc-fc measurements (see
text). Circles are obtained after applying field after cooling in zero field. Squares show results
of field-cooled measurements.}%
\label{fig2}%
\end{figure}

Figure \ref{fig1} shows typical magnetization loops for the cold - rolled sample (top
panel) and the most reversible sample of similar dimensions (bottom). The stressed sample
shows considerably more hysteresis than the stress-free sample. The hysteresis increases
for decreasing field in the stressed sample as expected from pinning, whereas the
hysteresis disappears approaching $H=0$ in the reversible sample, indicating complete
Meissner expulsion. In that sample pinning is also absent at larger fields as well. Also
shown in Fig.\ref{fig1} are MO images taken at the same temperature and magnetic fields
indicated by arrows. The flux structure for the stressed sample is dendritic and a
significant amount of flux is trapped at $H=0$. In the stress-free sample the patterns
are noticeably different, revealing flux tube phase upon flux penetration and
well-defined laminar pattern upon flux exit. This behavior was observed at all accessible
temperatures.

Figure \ref{fig2} shows details of the evolution of flux patterns in the stress-free
sample at $T=6$ K. After cooling in zero field, a full magnetization loop was measured
with a maximum magnetic field exceeding $H_c\approx 260$ Oe. Shown by the solid line, the
$M(H)$ loop exhibits magnetic hysteresis at intermediate fields. The crucial question is
whether this hysteresis is due to extrinsic factors (defects or residual stress) or it is
an intrinsic property of the intermediate state. To clarify this, we performed zero-field
and field - cooling experiments. The results are shown by the symbols in Fig.~\ref{fig2}.
The circles are obtained after cooling in zero field to 6 K and then increasing field,
whereas squares indicate measurements after cooling the sample to 6 K in a particular
field. If the hysteresis were due to pinning, the zero field cooled circles should
coincide with the ascending branch of the $M\left( H \right)$ loop but field-cooled
squares should not. Instead, Fig.\ref{fig2} clearly shows that both circles and squares
coincide exactly with the directly measured magnetization loop, implying that the
hysteresis is not due to pinning. The fact that the ascending and descending branches
merge at small and large fields is also inconsistent with pinning. Also, we observed no
magnetic relaxation with either flux penetration or flux exit. These results strongly
suggest that the hysteresis in stress-free samples is due to the topological difference
between the closed flux tube phase and the open laminar phase.

MO images shown in Fig.~\ref{fig2} reveal that after pure Meissner screening, the
intermediate state appears, not as laminae, but as an assembly of normal tubes carrying
magnetic flux and separated by superconducting regions. (MO image $\sharp 1$ shows the
entire sample, others zoom in to reveal the structure). These images show that flux tubes
have a variety of structures - from simple monodomain to complex objects threaded with
superconducting tubes. The flux tube phase favors hexagonal symmetry, almost exactly as
modeled by Goren and Tinkham \cite{15}. In all cases, this tubular phase has a closed
topology that allows screening currents to circulate \cite{4}. Similar patterns were
directly observed in In \cite{3,8}, Re \cite{3}, Sn \cite{3,21} and Hg \cite{4} and it
seems that closed topology tubular pattern is a generic feature of the intermediate state
of pinning-free type-I superconductors upon flux penetration.

Another possibility for the hysteresis is the edge barrier for flux penetration (including both,
Bean-Livingston and geometric barriers) \cite{17,18,19}. However, such a barrier would result in
delayed flux penetration and more negative values of magnetization compared to the thermodynamic
values. Figure \ref{fig2} shows that in our samples penetration occurs at the thermodynamic field
$H_c \left( {1-N} \right) \approx 120$ Oe and magnetization at that point is as supposed to be in
thermodynamic equilibrium, $4\pi M=-H_c $, independent of the demagnetization factor. We attribute
the weak influence of the edge barrier to the large thickness of our samples. On the other hand,
the edge barrier could be involved in the \emph{formation} of the observed topologies. When small
fingers of the normal phase are formed at the sample edge, the surface barrier will prevent their
continuous penetration into the interior and will break them into small flux tubes as suggested in
Ref.\cite{17}.

\begin{figure}[tb]
\centerline{\includegraphics[width=8cm]{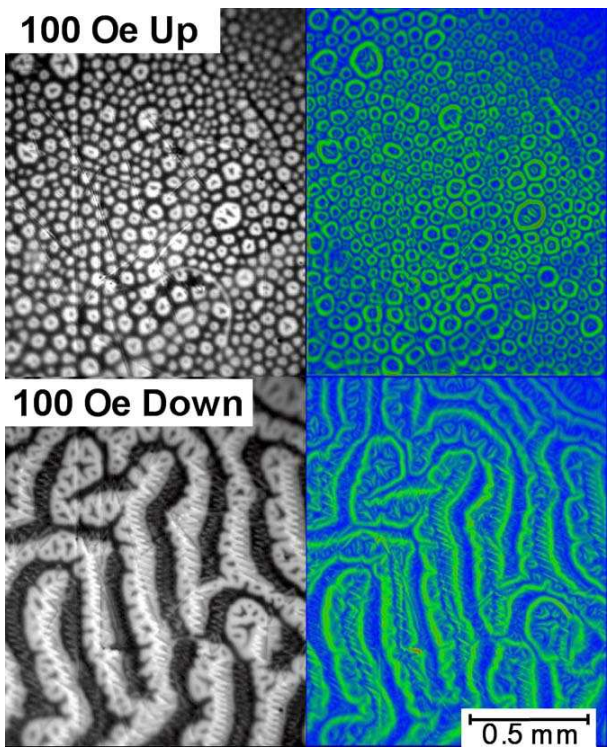}}%
\caption{$T=5$ K, $H=100$ Oe: (\textbf{left}) Distribution of the magnetic induction upon flux
penetration (top) and exit (bottom);(\textbf{right}) corresponding patterns of shielding currents
density obtained by numerical inversion. Intensity is proportional to the current density. (Color online)}%
\label{fig4}%
\end{figure}

Furthermore, quantitative imaging of the magnetic induction can be used to visualize
spatial distribution of shielding currents. This experimental information is important
for theoretical analysis involving current-loop models \cite{10,11} as well as for
general understanding of pattern formation in type-I superconductors. We used recently
developed fast inversion scheme \cite{perkins}. Figure \ref{fig4} shows the result
obtained for $H=100$ Oe. Left images correspond to flux penetration (top) and flux exit
(bottom) in our most reversible sample. The right panel shows corresponding distributions
of the shielding currents density (proportional to the brightness). Clearly, current
distribution exhibits two topologically distinct patterns. The closed topology of small
current loops vs. open topology of branched current streams.

\begin{figure}[tb]
\centerline{\includegraphics[width=8cm]{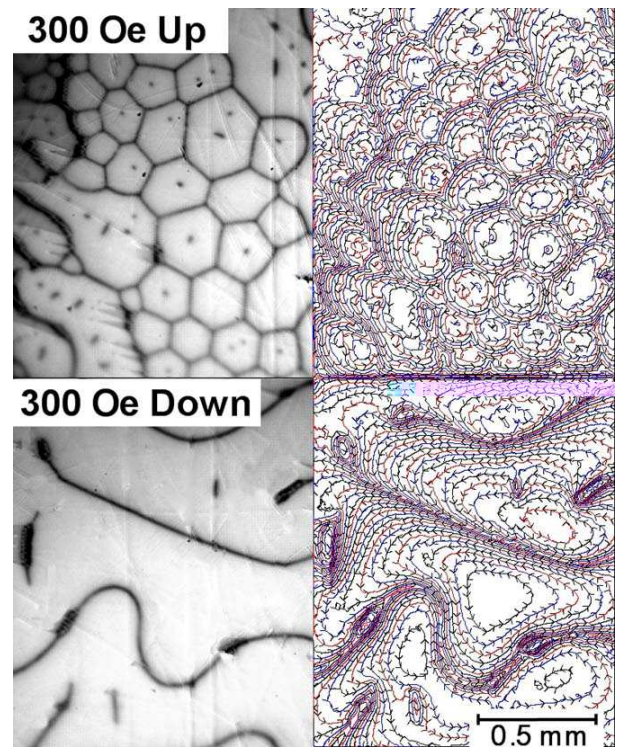}}%
\caption{$T=5$ K, $H=300$ Oe: (\textbf{left}) Distribution of the magnetic induction upon flux
penetration (top) and exit (bottom);(\textbf{right}) corresponding contour plots of shielding
currents density obtained by the numerical inversion. Arrows show the direction of currents. (Color online)}%
\label{fig5}%
\end{figure}

The direction of the currents flow is seen in Fig.\ref{fig5}, which shows the reconstruction at
$H=300$ Oe. At this field the features are larger and contour lines with directional arrows can be
used to better visualize the flow patterns. In both topologies, currents flow counterclockwise -
against the direction of Meissner currents flowing along the sample edges. At low fields Meissner
currents dominate and hysteresis is negligible. It appears only at intermediate fields when the
intermediate state consists of mobile ensemble of flux tubes on field entry and laminar structure
that forms escape paths for flux expelled by the Meissner effect upon flux exit.

\begin{figure}[tb]
\centerline{\includegraphics[width=7.5cm]{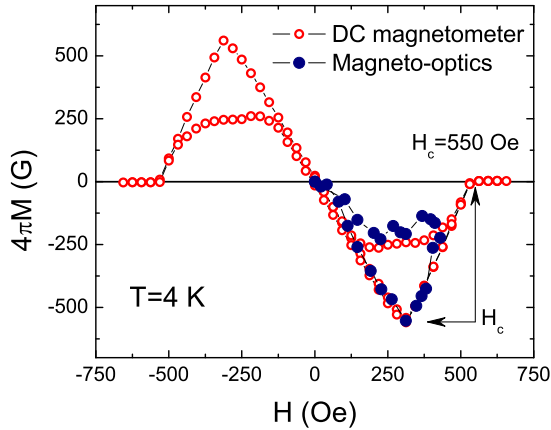}}%
\caption{Comparison of DC magnetization measured by magnetometer and reconstructed from MO images.
Arrows show that at the minimum $4\pi M=-H_c$ as expected at thermodynamic equilibrium
without edge barriers.}%
\label{fig3}%
\end{figure}

To evaluate quantitative correspondence of the MO and $M\left(H\right)$ measurements we calculate
total magnetic moment from the MO images by using $4\pi
M=\int{\left[\textbf{H}-\textbf{B}(\textbf{r})\right]}d^3\textbf{r}$. Magnetic induction is
linearly proportional to the intensity, hence integrating images and using initial slope of the
measured $M(H)$ loop for calibration, $M$ is obtained. Figure \ref{fig3} shows that $M(H)$ loop
from the MO images (solid symbols) is in a good agreement with the direct measurements (open
symbols).

Observed topological hysteresis is also clearly seen on the profiles of the magnetic
induction. Following Landau \cite{2}, it has generally been assumed that the magnetic
field inside the normal phase in the intermediate state is close to $H_{c }$
\cite{3,4,9,17}. Figure \ref{fig3} shows profiles of the magnetic induction measured at
the same external field in the flux tube phase (obtained on flux entry) and in a laminar
phase (obtained on flux exit). While the field in the laminae is comparable to the
critical field $H_{c}$ as expected, the field above the flux tubes is much smaller.
Indeed, the measurements are carried out $\approx 10 \mu$m above the sample, so the
measured field is reduced compared to the values inside the tubes. However, simple
numerical analysis with appropriate dimensions does not reproduce such substantial
reduction. It is possible that flux tubes widen when approaching the surface \cite{3},
but why it is not seen in the laminar phase?

\begin{figure}[tb]
\centerline{\includegraphics[width=8cm]{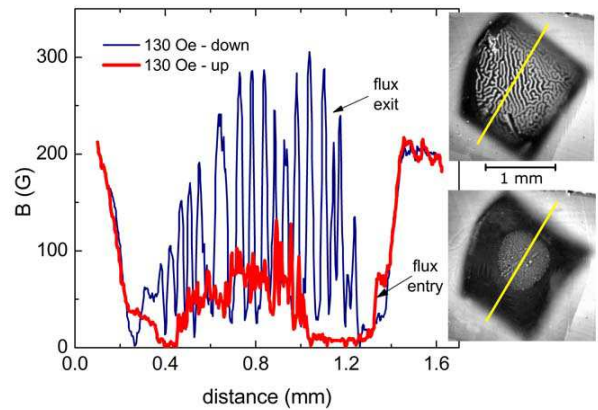}}%
\caption{Profiles of the magnetic induction measured at $T=4$ K in the flux tube phase
(flux penetration)and laminar phase (flux exit). Right panel shows corresponding
MO images.}%
\label{fig6}%
\end{figure}

Alternatively, it is possible that when a flux tube initially appears at the sample edge
with critical magnetic field inside. Due to closed topology, the total magnetic flux in
such tube is now conserved. The nucleated flux tube is driven by Meissner currents toward
sample interior (these currents flow everywhere on the surfaces perpendicular to the
magnetic field \cite{20}) until it is stopped at the center or later by other tubes
piling up from the center outward. When the flux tube reaches the interior, its radius
may increase to minimize the magnetic field energy. Real time imaging showed that flux
tubes produced at the sample edge continue to travel toward the center and form an
apparently outwardly expanding phase \cite{21}, as seen in the lower-right panel of
Fig.\ref{fig6}. The real-time observations were first made by Solomon and Harris in 1971
\cite{21}. The rigorous evaluation of the free energy of even a single tube is not
simple. The difficulty is the lack of a sharp interface between the superconductor and
tube interior, which decreases the surface energy. The tubes repel each other at the
large distances due to interaction of screening currents, but they attract each other
when their "cores" overlap. When tubes merge, average magnetic field in the tube
increases until it reaches $H_c$. At this stage, the honeycomb lattice is formed and it
persists almost up to $H=H_c$. The observed topological hysteresis is observed only at
the stage when magnetic field inside the tubes is less than $H_c$ and before the
formation of rigid hexagonal lattice, Fig.\ref{fig2}. A detailed study of tube nucleation
and expansion is needed to quantify this issue.

We thank J.~R.~Clem, A.~T.~Dorsey, T.~A.~Girard, N.~D.~Goldenfeld, R.~P.~Huebener,
R.~V.~Kohn, D.~C.~Larbalestier and V.~K.~Vlasko-Vlasov for useful discussions.

\end{document}